\documentclass[sigconf]{aamas}

\usepackage{balance} 
\balance

\usepackage{xcolor}

\usepackage[linesnumbered,ruled,vlined]{algorithm2e}

\SetCommentSty{mycommfont}

\usepackage{amsmath}

\usepackage{hyperref}
\usepackage{tcolorbox}
\usepackage{listings}

\usepackage[inline]{enumitem}

\usepackage{xparse} 
\usepackage[textsize=small]{todonotes}
\let\oldtodo\todo
\RenewDocumentCommand{\todo}{o m}{%
  \IfNoValueTF{#1}
    {\oldtodo[inline]{#2}}
    {\oldtodo[inline,#1]{#2}}
}

\doi{HKPK4104}

\makeatletter
\gdef\@copyrightpermission{
  \begin{minipage}{0.2\columnwidth}
   \href{https://creativecommons.org/licenses/by/4.0/}{\includegraphics[width=0.90\textwidth]{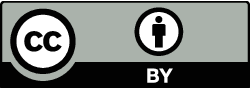}}
  \end{minipage}\hfill
  \begin{minipage}{0.8\columnwidth}
   \href{https://creativecommons.org/licenses/by/4.0/}{This work is licensed under a Creative Commons Attribution International 4.0 License.}
  \end{minipage}
  \vspace{5pt}
}
\makeatother

\setcopyright{ifaamas}
\acmConference[AAMAS '26]{Proc.\@ of the 25th International Conference
on Autonomous Agents and Multiagent Systems (AAMAS 2026)}{May 25 -- 29, 2026}
{Paphos, Cyprus}{C.~Amato, L.~Dennis, V.~Mascardi, J.~Thangarajah (eds.)}
\copyrightyear{2026}
\acmYear{2026}
\acmDOI{}
\acmPrice{}
\acmISBN{}

\acmSubmissionID{1194}

\title{Self-Evolving Software Agents}
\subtitle{Extended Abstract}

\author{Marco Robol}
\affiliation{
  \institution{University of Trento}
  \city{Trento}
  \country{Italy}
}
\email{marco.robol@unint.it}

\author{Paolo Giorgini}
\affiliation{
  \institution{University of Trento}
  \city{Trento}
  \country{Italy}
}
\email{paolo.giorgini@unitn.it}

\begin{abstract}

Autonomous agents can adapt their behaviour to changing environments, but remain bound to requirements, goals, and capabilities fixed at design time, preventing genuine software evolution. This paper introduces \emph{self-evolving software agents}, combining BDI reasoning with LLMs to enable autonomous evolution of goals, reasoning, and executable code.
We propose a BDI--LLM architecture in which an automated evolution module operates alongside the agent’s reasoning loop, eliciting new requirements from experience and synthesizing corresponding design and code updates. A prototype evaluated in a dynamic multi-agent environment shows that agents can autonomously discover new goals and generate executable behaviours from minimal prior knowledge.
The results indicate both the feasibility and current limits of LLM-driven evolution, particularly in terms of behavioural inheritance and stability.

\end{abstract}

\ccsdesc[500]{Computing methodologies~Artificial intelligence}
\ccsdesc[100]{Computing methodologies~Philosophical/theoretical foundations of artificial intelligence}
\ccsdesc[100]{Computing methodologies~Intelligent agents}
\ccsdesc[300]{Computing methodologies~Generative and developmental approaches}
\ccsdesc[100]{Computing methodologies~Simulation evaluation}
\ccsdesc[300]{Software and its engineering~Software creation and management}
\ccsdesc[300]{Software and its engineering~Software organization and properties}

\keywords{Software Evolution; Adaptation; Autonomous Agents; BDI model; Artificial Intelligence; LLMs}

\newcommand{\BibTeX}{\rm B\kern-.05em{\sc i\kern-.025em b}\kern-.08em\TeX}

\begin{document}


\pagestyle{fancy}
\fancyhead{}


\maketitle 


\section{Motivation and Background}

Modern software systems increasingly operate in open and dynamic environments shaped by machine learning, IoT, and cloud computing~\cite{paris2021, bettini2015}, where requirements, assumptions, and operational contexts evolve over time~\cite{bohm2020, davenport2019}. While self-adaptive systems and autonomous agents can modify their behaviour at runtime, they typically remain bound to goals, requirements, and capabilities defined at design time. As a result, they can adapt, but not genuinely evolve.
Software evolution, originally limited to code fixes and incremental updates~\cite{lehman1980}, now entails revising system objectives, internal reasoning structures, and executable capabilities in response to emerging needs~\cite{cheng2009}. Despite extensive research on software evolution~\cite{Fernandez-Ramil06, swchange1999, somerville2010, pressman2005}, requirements engineering~\cite{parnas1994, boehm1988, jackson1995}, and self-managing systems~\cite{deLemos2001, oreizy1999, garlan2004}, current agent architectures lack explicit mechanisms to autonomously evolve their own requirements and code while preserving architectural coherence~\cite{mcKinley2004, swadaptation1997}.
Recent advances in Large Language Models~\cite{bommasani2021} offer new opportunities to automate parts of the software evolution process, including requirement elicitation, design revision, and code synthesis. However, existing agentic AI approaches largely rely on prompt-driven behaviour and externally defined objectives, providing limited support for structured, long-term evolution.

This work addresses this gap by introducing a framework for self-evolving software agents that integrates automated software evolution principles within a BDI architecture, enabling agents to evolve goals, reasoning, and actions autonomously. 

\section{Self-Evolving Agents Architecture}
\begin{figure*}[!htbp]
    \centering
    \includegraphics[width=0.9\textwidth]{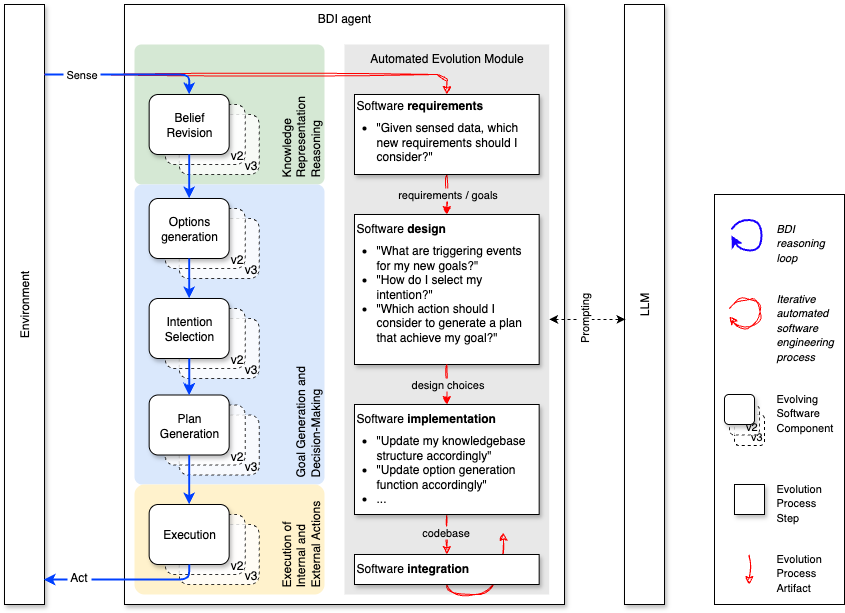}
    \Description{Diagram showing the architecture of a self-evolving BDI agent.}
    \caption{BDI--LLM architecture for self-evolving software agents. An automated evolution module operates alongside the classical BDI reasoning loop, enabling autonomous evolution of knowledge, goals, and executable actions.}
    \label{fig:self-evolvingBDI}
\end{figure*}

Figure~\ref{fig:self-evolvingBDI} illustrates the proposed BDI--LLM architecture for self-evolving software agents~\cite{franklin1996, wooldridge2009, luck2021agent}. The model builds on a classical BDI reasoning loop~\cite{rao1995, wooldridge1995, wooldridge2009}, where the agent continuously updates beliefs, deliberates over desires, selects intentions, and executes plans in response to environmental perceptions.
In addition to the standard reasoning loop, the architecture introduces an \emph{Automated Evolution Module} that operates independently from runtime decision making. This module monitors the agent’s experience and identifies unmet needs or novel opportunities that cannot be addressed by the current knowledge, goals, or action repertoire. When triggered, it initiates an automated software evolution cycle based on variation, selection, and inheritance.

The evolution process acts on three architectural layers of the agent: (i) knowledge representation and reasoning, by extending or revising perceptual and inference mechanisms; (ii) goal generation and decision making, by introducing new goals and intention-selection policies; and (iii) execution, by synthesizing or adapting executable actions and plans. By isolating evolution from the reasoning loop, the architecture preserves behavioural coherence while enabling autonomous, long-term evolution of the agent's internal structure. This approach addresses challenges in adaptive systems~\cite{muller2014application, whittle2011, cheng2009} and leverages recent advances in LLMs for automated code generation~\cite{brown2020language, chen2021evaluating, bommasani2021, li2022alphacode}.

\section{Prototype and Preliminary Evaluation}

We implemented a prototype of the proposed self-evolving agent using a BDI control loop extended with an LLM-driven evolution module~\cite{openai2024gpt4o, vaccari2023}. The agent operates in a dynamic multi-agent environment inspired by the Deliveroo.js framework, where it must perceive the environment, discover goals, and generate executable plans while interacting with other agents and environmental constraints. The agent is initially provided only with a textual description of the environment and a minimal set of APIs, without predefined domain-specific knowledge or goals.
The evolution module is triggered when perceived information cannot be interpreted or exploited using the current knowledge and goal structures. In such cases, the agent autonomously generates new goals, revises its reasoning structures, and synthesizes executable actions, which are then validated through interaction with the environment. Successful behaviours are retained and reused in subsequent situations, while ineffective ones are discarded.

Preliminary experiments show that the agent is able to au\-to\-no\-mous\-ly discover operational goals and generate executable behaviours starting from minimal prior knowledge. At the same time, the results highlight current limitations of LLM-driven evolution, particularly in terms of behavioural inheritance and robustness as environmental complexity increases. These observations confirm the feasibility of autonomous software evolution while motivating further investigation on mechanisms for stabilising and reinforcing evolved behaviours. 

%
%
\section{Conclusions and Outlook}
This paper introduced a framework for self-evolving software agents that integrates automated software evolution principles within a BDI architecture augmented by LLMs. By separating runtime reasoning from an explicit evolution module, the proposed approach enables agents to autonomously revise goals, reasoning structures, and executable actions, moving beyond traditional notions of behavioural adaptation.
The prototype and preliminary evaluation demonstrate the feasibility of autonomous evolution in dynamic multi-agent environments, while also revealing current limitations in behavioural inheritance, stability, and scalability. These challenges point to the need for reinforcement mechanisms, memory consolidation, and more robust selection strategies.

Future work will focus on strengthening inheritance and long-term consistency, extending the evaluation to more complex multi-agent scenarios, and exploring collective and cooperative forms of software evolution among agents, as well as leveraging retrieval-augmented generation~\cite{lewis2020rag} to enhance LLM reasoning with external knowledge. 



\bibliographystyle{ACM-Reference-Format} 
\bibliography{bib}



\newpage

\color{red}





\color{black}


\end{document}